\documentclass[twocolumn,aps]{revtex4}

\usepackage{epsfig}
\usepackage{amsmath,amssymb,color}
\usepackage{footnote}
\usepackage[english]{babel}

\parskip=\medskipamount

\newcommand{\eq}[1]{(\ref{#1})}
\newcommand{\fig}[1]{Fig.\ref{#1}}

\newcommand{\be}{\begin{equation}}
\newcommand{\ee}{\end{equation}}

\newcommand{\la}{\langle}
\newcommand{\ra}{\rangle}

\begin{document}

\title{Self-isolation or borders closing: what prevents epidemic spreading better?}

\author{O. Valba$^{1,2}$, V. Avetisov$^2$, A. Gorsky$^{3,4}$, and S. Nechaev$^{5,6}$}

\affiliation{$^1$Department of Applied Mathematics, National Research University Higher School of Economics, 101000, Moscow, Russia \\ $^2$Federal Research Center of Chemical Physics RAS, 119991, Moscow, Russia \\ $^3$Institute of Information Transmission Problems RAS, 127051 Moscow, Russia \\ $^4$Moscow Institute of Physics and Technology, Dolgoprudny 141700, Russia \\ $^5$Interdisciplinary Scientific Center Poncelet, CNRS UMI 2615, 119002 Moscow, Russia \\ $^6$P.N. Lebedev Physical Institute RAS, 119991 Moscow, Russia}

\begin{abstract}

Pandemic distribution of COVID-19 in the world has motivated us to discuss combined effects of network clustering and adaptivity on epidemic spreading. We address the question concerning the choice of optimal mechanism for most effective prohibiting disease propagation in a connected network: adaptive clustering, which mimics self-isolation (SI) in local communities, or sharp instant clustering, which looks like frontiers closing (FC) between cities and countries. SI-networks are "adaptively grown" under condition of maximization of small cliques in the entire network, while FC-networks are "instantly created".  Running the standard SIR model on clustered SI- and FC-networks, we demonstrate that the adaptive network clustering prohibits the epidemic spreading better than the instant clustering in the network with similar parameters. We found that SI model has scale-free property for degree distribution $P(k)\sim k^{\eta}$ with small critical exponent $-2<\eta<-1$ and argue that scale-free behavior emerges due to the randomness in the initial degree distributions and is absent for random regular graphs.

\end{abstract}

\maketitle

\section{Introduction}

It is known \cite{review} that any epidemic spreading is sensitive to two generic features: clustering and adaptivity. Both of them have a strong impact on epidemic threshold \cite{clust1,clust2, clust3,adapt1}, peak value and typical distribution time. Here we are focused on a specific mechanism of adaptive clustering, which has strong impact on the disease propagation. Our work, inspired by the pandemic distribution of COVID-19 in the world, motivated us to discuss combined effects of network clustering and adaptivity on epidemic spreading.

We are encouraged by an observation made in \cite{anderson} concerning localization of one-body excitations on network clusters obtained in a specific evolutionary way. More recently, similar results have been derived for networks with different patterns of dynamically induced clustering \cite{loc1, loc2,loc3}. In the current Letter we analyze and compare numerically the epidemic spreading on adaptively and instantly clustered connected networks.

In \cite{anderson} we have considered spectral properties of two types constrained random Erd\H{o}s-R\'{e}nyi networks in the clustered phase: (i) "e-networks" obtained by the evolutionary Metropolis maximization of small cliques, and (ii) "i-networks", instantly prepared clustered graphs having the same geometrical properties as "e-networks", but which are created without any evolutionary selection. In e-networks, which are non-ergodic, excitations are mostly localized on clusters and weakly spread through the entire network, since the network is connected and there is a small, though finite density of inter-cluster links. Ergodic i-networks, which serve as a particular example of a "stochastic block model" \cite{fortunato10, decelle13}, being geometrically very similar to e-networks, are less effective in blocking the excitation spreading. As we show below, the distinction between e- and i-networks deals with different statistics of inter-cluster links in these networks. In our work we report results of simulations of the standard SIR model on clustered e- and i-networks. The SIR model (described in Section IV) is the simplest and widely used model of disease transmission from human to human.

The paper is structured as follows. In Section II we formulate the model of adaptive clustering. In Section III we argue that our model has a scale-free degree distribution providing explanation of a very specific triangular shape of the spectral density of clustered e-networks observed in \cite{hovan}. In Section IV we describe results of simulations of SIR model on Erd\H{o}s-R\'{e}nyi (ER), e- and i-networks. In Discussion we speculate about possible interpretation of self-isolation (SI) in communities as formation of adaptively clustered e-networks, frontiers closing (FC) -- as formation of i-networks, and demonstrate that SI prohibits the epidemic spread more efficient than FC.

\section{Definitions and networks generation}

The main object of our consideration is the dynamically evolving constrained Erd\H{o}s-R\'enyi network. The $N$-vertex Erd\H{o}s-R\'enyi network is a topological graph of $N$ vertices constructed by random linking with probability $p$ any pair of points from a set of $N$ arbitrary points. The probability, $P(k)$, to find a vertex in ER network, linked with other $k$ vertices, is Poissonian with the mean value $\la k\ra = Np$. Another well-studied class of random networks are the so-called scale-free networks, for which the vertex degree distribution, $P(k)\sim k^{\eta}$, has a power-law tail with a critical exponent $\eta<0$ (typically $\eta<-2$). The overwhelming majority of natural networks is scale-free, and the network of distribution of COVID-19 is not an exception \cite{vir}.

Natural networks, being complex self-organized objects, evolve in time trying to adapt themselves to imposed external conditions. We distinguish two classes of dynamic Erd\H{o}s-R\'enyi networks: "unconstrained" (without the vertex degree conservation during the network evolution) and "constrained" (with preservation of vertex degrees in all nodes under network rewiring). In unconstrained ER networks one can remove any link from one place of the network and insert it into any other place. To the contrary, in constrained ER networks the realization of a rewiring is more complex and involves simultaneous replacement at least two bonds.

Speaking less abstract, consider a network of human social relations, where each graph vertex represents a particular individual. It seems reasonable to assume that for each individual, the number of social connections (the particular vertex degree in a social network) is conserved. The number of connections may vary from one individual to another, however for each human it is supposed to be fixed and unchanged during the social network evolution. Such a supposition seems rather natural since the number of relations per one individual rapidly increases, saturates and then remains approximately conserved in time. Specifically, we proceed with the following rewiring setup which conserves vertex degrees. We take a random Erd\H{o}s-R\'enyi $N$-vertex graph without double connections as an initial state of a network. Then, we randomly select a pair of arbitrary links, say, $(ij)$ (between vertices $i$ and $j$) and $(k,l)$ (between $k$ and $l$), and reconnect them, getting new links $(i,k)$ and $(j,l)$. Such reconnections conserve the vertex degree \cite{maslov}, however allow for bonds redistribution and do not prohibit topological changes in the entire network. In the context of phase transitions in social networks such dynamic model has been discussed in \cite{schelling}.

The following question has been addressed in \cite{hovan}. Suppose that we rewire links in the constrained Erd\H{o}s-R\'enyi network under the condition that at each step of rewiring we try to maximize the number of small cliques (small complete subgraphs of few links). Which is the equilibrium structure of the entire network? In mathematical terms this question reads as follows. We assign the energy $\mu$ to each simplest clique (closed triad of bonds) and denote by $n_{\triangle}$ the number of such triads in the network. The partition function of the network can be written as
\be
Z(\mu) = \sum_{\{\rm states\}} \hspace{-0.25cm} {\vphantom{\sum}}' e^{-\mu n_{\triangle}}
\label{eq:01}
\ee
where prime in \eq{eq:01} means that the summation runs over all possible configurations of
links ("states"), under the condition of fixed degrees $\{v_1,...,v_N\}$ in all network vertices.

To simulate the rewiring process, one applies the standard Metropolis algorithm with the following rules: i) if under the reconnection the number of closed triads is increasing, a move (rewiring)
is accepted, ii) if the number of closed triads is decreasing by $\Delta\, n_{\triangle}$, or remains unchanged, a move is accepted with the probability $e^{-\mu \Delta\, n_{\triangle}}$. The Metropolis algorithm runs repeatedly for large set of randomly chosen pairs of links, until it converges. In \cite{reconnection} it was proven that such Metropolis algorithm converges to the Gibbs measure $e^{\mu N_{\triangle}}$ in the equilibrium ensemble of random undirected Erd\H{o}s-Renyi networks with fixed vertex degree.

In \cite{hovan} it has been shown that given the bond formation probability, $p$, in the initial graph, the evolving network splits into the maximally possible number of clusters, $N_{cl}$:
\be
N_{cl}=\left.\left [\frac{N}{Np+1} \right]\right|_{N\gg 1}\approx \left [\frac{1}{p}\right],
\label{eq:02}
\ee
where $[x]$ means the integer part of $x$ and the denominator $(Np+1)$ defines the minimal size of
formed cliques. The asymptotic limit $\sim [p^{-1}]$ at $N\to\infty$ in \eq{eq:02} is independent on the particular set of corresponding vertex degrees, $\{v_1,...,v_N\}$.

\begin{figure}[ht]
\centerline{\includegraphics[width=8cm]{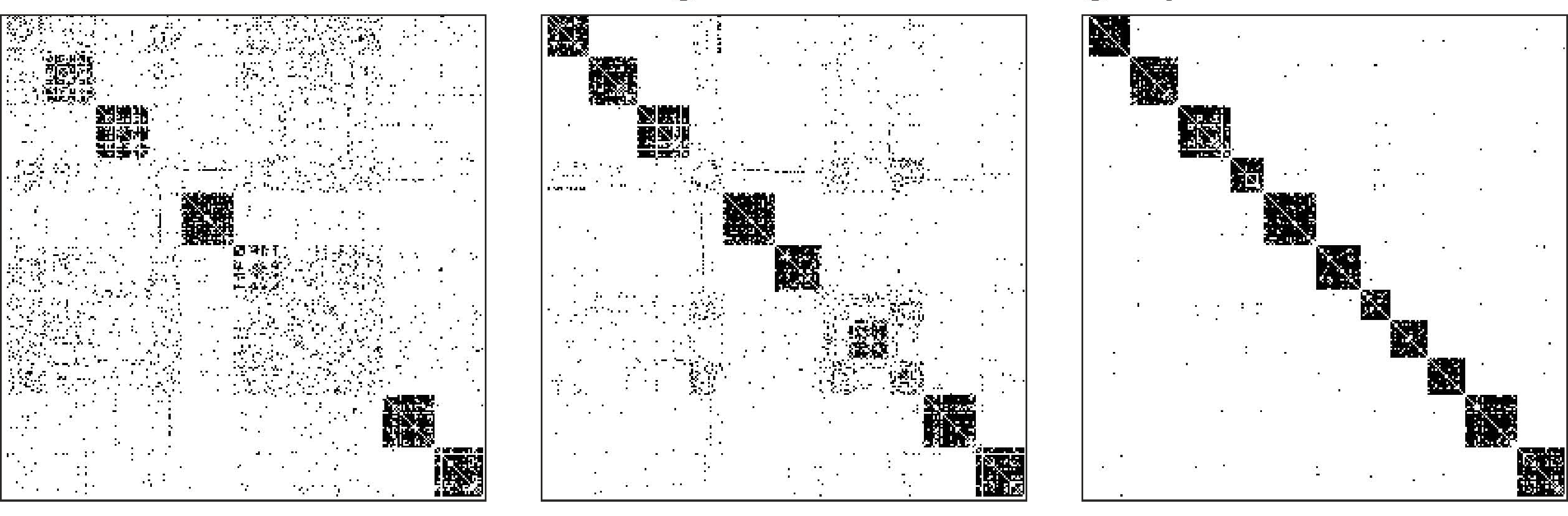}}
\caption{Few typical samples of intermediate stages of a network evolution at fixed vertex degree under condition of triads maximization.}
\label{fig:01}
\end{figure}

According to the work \cite{hovan}, clustering of evolving constrained Erd\H{o}s-Renyi network occurs under condition of triads maximization, as a first order phase transition where $\mu$ is a control value. To have some insight about topological network structure in course of its evolution under maximization of triadic motifs, we reproduce in \fig{fig:01} typical adjacency matrices at three sequential stages of a particular network rearrangement -- see \cite{hovan} for details.

To visualize the evolution, we enumerate vertices at the preparation condition in arbitrary order and run the Metropolis stochastic dynamics. When the system is equilibrated and clusters are formed, we re-enumerate vertices sequentially according to their belongings to clusters. Then we restore corresponding dynamic pathways back to the initial configuration.

\begin{figure}[ht]
\centerline{\includegraphics[width=8cm]{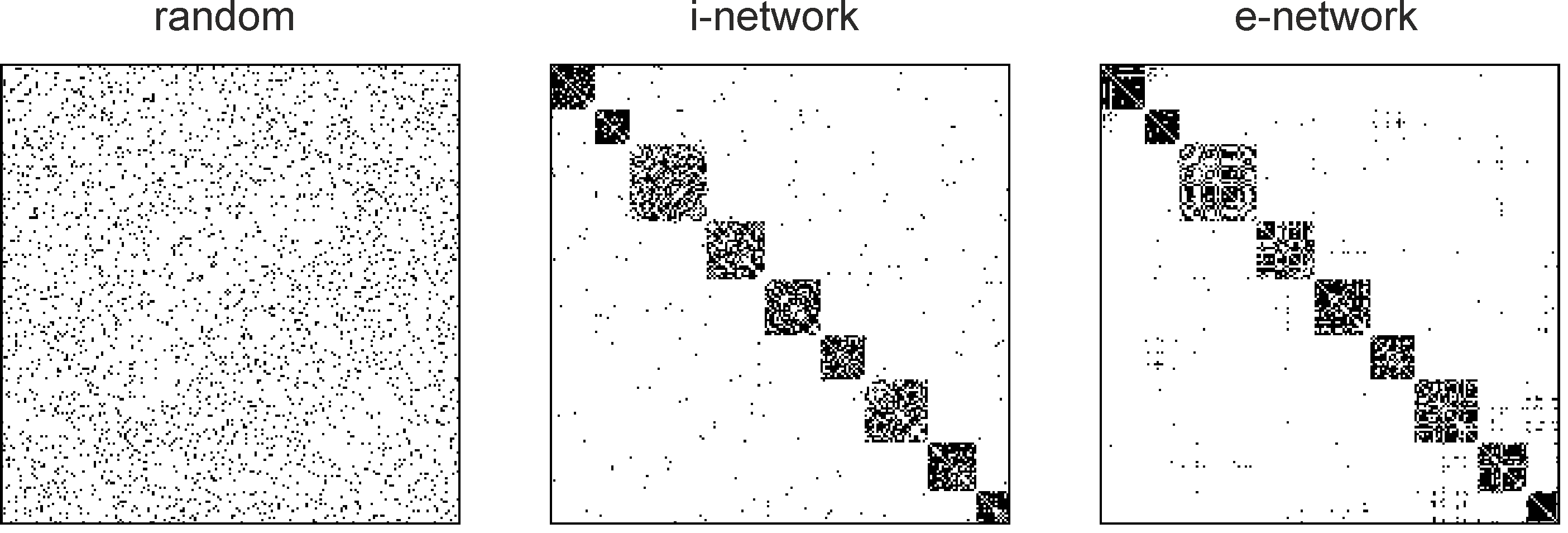}}
\caption{Examples of typical adjacency matrices for random, e- and i-networks, which have $N=750$ vertices and are created with the link probability $p=0.08$. By visual inspection it is almost impossible to distinguish the adjacency matrices of e- and i-networks.}
\label{fig:02}
\end{figure}

The evolutionary grown clustered "e-network", obtained by the maximization of triangles (triadic motifs) we compare with another mechanism of clustered "i-network" formation. The "i-network" is instantly formed being a particular example of a stochastic block random graph \cite{fortunato10, decelle13}. The i-network is constructed by the following procedure. Firstly, we detect clusters $\{J\}$ in the e-network, define the link probability, $p_{in}^{J}$, inside each cluster $J$, and between clusters, $p_{out}$. Secondly, we split the set of $N$ points in groups as in the e-network, and generate new random clusters by connecting vertices in these groups with the probability $p_{in}^{J}$. In such a way we "mimic" clusters of e-networks. Finally, we randomly connect nodes belonging to different clusters with the probability $p_{out}$ borrowed from the average connection probability between clusters in e-network. Such an "instantly created" i-network mimics the e-network, since i-network has the same linking probability and community structure as the evolutionary grown e-network -- see \fig{fig:02}. However the i-network has no any pre-history, it knows nothing about the evolution, it has no dependence on $\mu$, and has no vertex degree conservation. Visual inspection of \fig{fig:02} does not allow us to distinguish adjacency matrices e- and i-networks. Besides, propagation of excitation on e- and i-networks behaves very differently.

To summarize, we have in hands ensembles of two kinds of networks: (i) evolutionary grown (e-networks) which have memory about the history of its creation, and (ii) instantly \emph{ad hoc} formed (i-networks). Comparing mechanisms of construction of e- and i-networks, it seems plausible:
\begin{itemize}
\item[-] To identify clustered e-networks, obtained by adaptive preferential arrangement of network vertices in small cliques with self-isolation (SI) of humans in small communities,
\item[-] To identify instantly created i-networks with splitting of entire human network into collection of weakly connected clusters obtained by frontiers (borders) closing (FC).
\end{itemize}
For comparison we also consider random Erd\H{o}s-Renyi networks of the same vertex degree distribution, which are initial states of our evolutionary algorithm.

\section{Adaptive clustering and scale-free distribution in e-networks}

In \cite{hovan} we have pointed out some puzzling property of the spectral density (eigenvalue distribution) of adaptively clustered networks. The spectrum above clustering transition has two-band structure in which the first (main) band was naturally attributed with perturbative excitations inside clusters, while the second "non-perturbative" band emerged from eigenvalues tunneled from the first zone aside. It was found numerically that the spectral density in the perturbative band has triangular shape typical for scale-free networks \cite{scale1,scale2}. Such result looked surprising since the clustered network is originated from a standard Erd\H{o}s-R\'{e}nyi graph with a binomial degree distribution and since the vertex degree is conserved in the network evolution. Naively thinking there is no place for a network to be scale-free.

\begin{figure}[ht]
\centerline{\includegraphics[width=8cm]{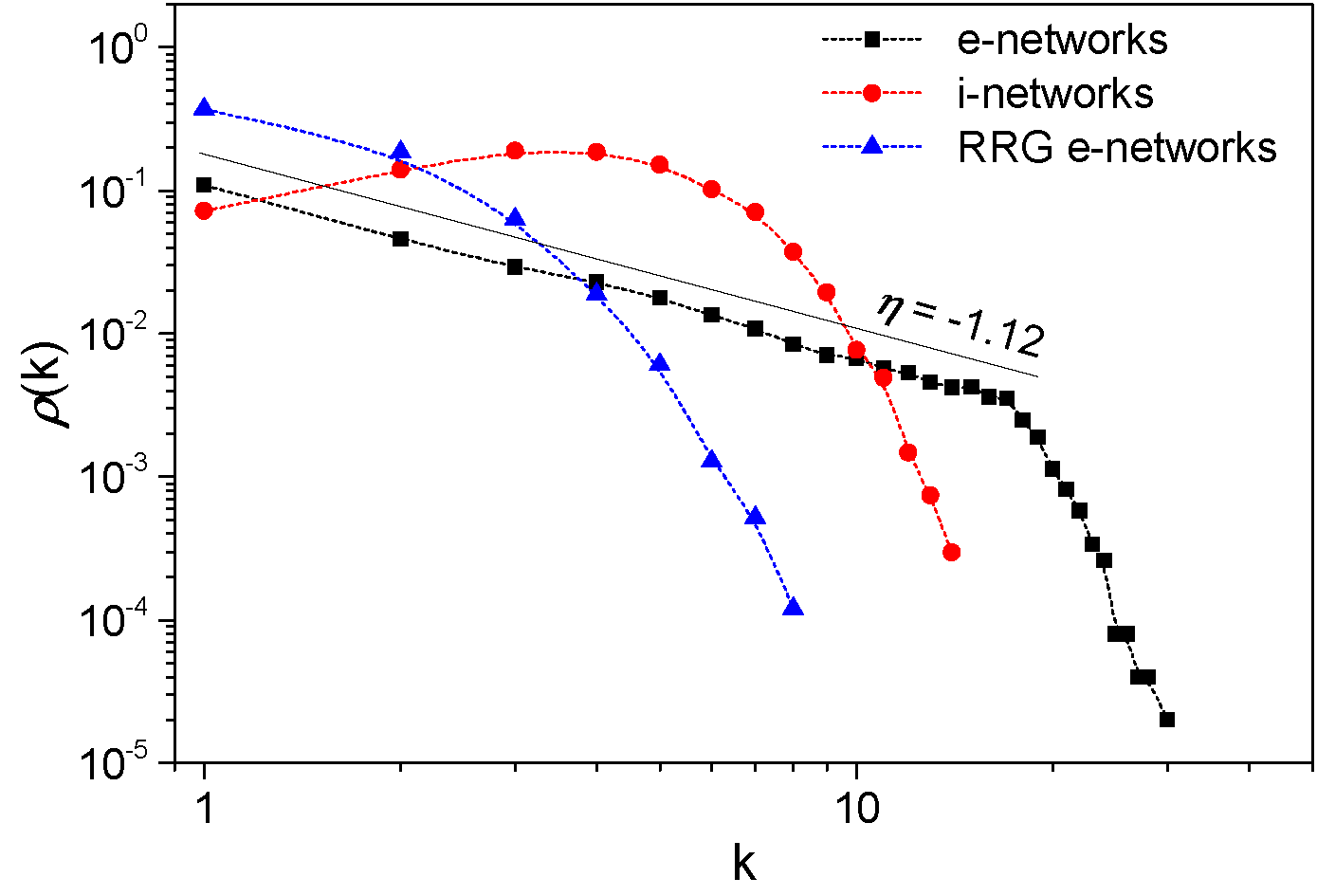}}
\caption{(a) The outer cluster degree distribution $\rho(k)$ in log-log scale. Results are obtained for 100 realizations of e-networks with $N=750$ nodes and the linking probability $p=0.08$; (b) The distributions for the constrained ER and random regular graphs (RRG).}
\label{fig:03}
\end{figure}

The resolution of that puzzle turns out to be as follows: we have to study separately  distributions of internal (inside cluster) and external (between clusters) vertex degrees. Consider a vertex $i$, which belongs to the cluster $J$ of clustered e-network, and define the "outer degree" for a vertex $i$ as the number of links, connecting $i$ to vertices of clusters different from $J$. In \fig{fig:03} we have plotted the "outer vertex degree distribution", $\rho(k)$, of cluster nodes for three types of networks: e-networks (black squares), i-networks (red circles) and RRG e-networks (blue triangles). Simulations show for e-networks the power-law scaling
\be
\rho(k)\sim k^{\eta}
\label{eq:04}
\ee
with surprisingly small value of $\eta$. The line of the best fit in \fig{fig:03} for "outer vertex degree" of e-networks has the slope $\eta=-1.12$. The "inner vertex degree distribution" demonstrates at the same time the binomial distribution modified by the long tail at small degrees. Instantly created i-networks (red circles) do not possess such scale-free behavior for vertex degrees between clusters.

It is eligible to ask a question which property of e-network is responsible for the scale-free distribution. To this aim we consider the e-networks constructed on the basis of random regular graphs (RRG), possessing the similar cluster structure. The outer vertex degree distribution of RRG e-networks is shown by blue triangles in \fig{fig:03} and demonstrates the absence of scale-free behavior, however the distribution itself seems to be closer to the one of e-networks, rather than of i-networks. Thus, we have solid arguments to suggest that the scale-free behavior of e-networks is induced by the disorder in the vertex degree distribution of the "parent" constrained ER network.

The dependence \eq{eq:04} is fully consistent with our investigations \cite{hovan, anderson} of spectral statistics of evolutionary grown clustered networks. It was shown in \cite{hovan} that the enveloping shape of the main band in spectral density of the adjacency matrix is changing with increasing of $\mu$ from the semicircle (in the initial ER network and below $\mu_{cr}$) to the triangle (above $\mu_{cr}$, in the clustered network), where $\mu_c$ is first order transition point. According to our observation, the triangular shape of the spectral density in the main band should be attributed mainly to the scale-free property of inter-cluster excitations of e-networks.

The critical exponent $\eta$ is small ($-2<\eta<-1$), which means that the average vertex degree distribution diverges. The general conditions to have $\eta>-2$ in scale-free networks with fixed number of nodes is discussed in \cite{bianconi,dorog}. Rather nontrivial rewiring procedure for generating networks with $\eta>-2$ has been proposed in \cite{bianconi}. Despite the algorithm of \cite{bianconi} looks rather sophisticated, the ideas behind its construction are in good agreement with our simple generation procedure of the scale-free network with $-2<\eta<-1$. Namely, to get $\eta>-2$ one should accurately tune the combination of local and global constraints. One more rewiring procedure was suggested in \cite{dorog} for getting $\eta=-1$. Fortunately, our simple algorithm dealing with maximization of triads in constrained Erd\H{o}s-R\'{e}nyi network brings the system automatically in the regime where generation of a scale-free subnetwork with $-2<\eta<-1$ occurs. To summarize, the main result of this section consists in providing a new simple rewiring mechanism for the fabrication of scale-free distribution in constrained Erd\H{o}s-R\'{e}nyi networks.

\section{Numerical simulation of SIR model}

Epidemic models classify individual agents (humans) based on the stage of disease affecting them. The simplest classification scheme assumes that an individual can be in one of three states (compartments): (a) susceptible (S) for healthy individuals having not yet contacted the pathogen, (b) infectious (I) for contagious individuals have contacted the pathogen and can infect others, (c) recovered (R) for recovered (or immune) individuals. The distribution of disease on some target space is considered in the frameworks of transformation between susceptible, infectious and recovered agents and is known as the SIR model \cite{sir}. The standard dynamics of SIR model reads:
\be
\begin{aligned}
S+I & \xrightarrow{\beta} I+I \\
I & \xrightarrow{\gamma} R
\end{aligned}
\label{eq:03}
\ee
The model has two adjustable parameters $(\beta, \gamma)$. These parameters set transition rates, $\beta$, for susceptible nodes to become infected from infected neighbors, and $\gamma$, for infected nodes to recover.

\begin{figure}[ht]
\centerline{\includegraphics[width=9.5cm]{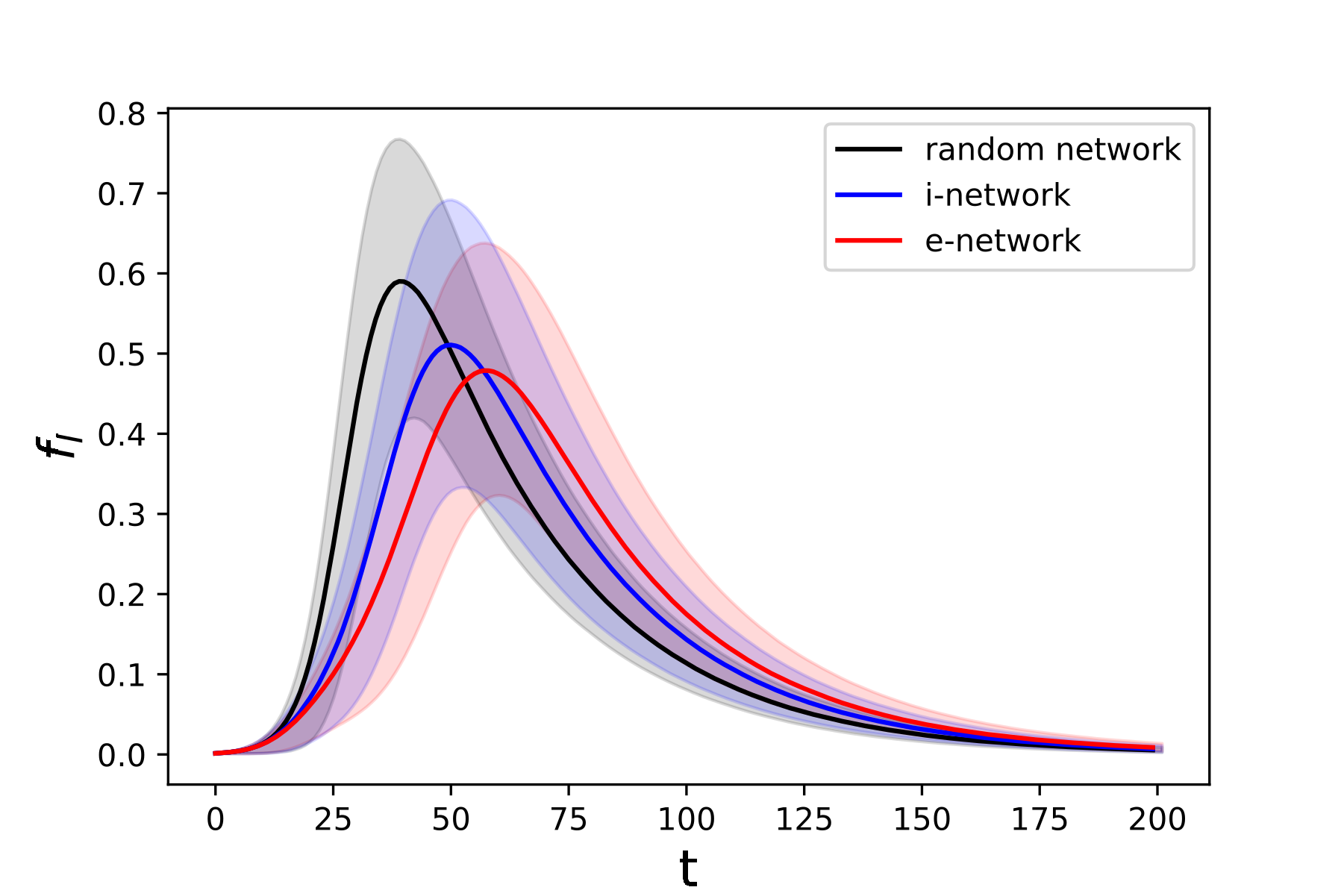}}
\caption{The fraction of infected nodes in time for random, i- and e-networks. Results were obtained for the SIR dynamics  with transmission rates $\beta=0.05$ and recovery rate $\gamma=0.03$ and with $n = 1000$ simulations on each networks. Shadowed regions designate the confidence range of respective network archetype.}
\label{fig:04}
\end{figure}

We have run the SIR model on three archetypes of graphs: random Erd\H{o}s-Renyi (ER) network, e-network and i-network, the respective adjacency matrices are shown in \fig{fig:02}. The results of our simulations are depicted in \fig{fig:04}, where we have plotted the density of infected agents, $f_i$, versus time, $t$. To be able to compare distributions, we developed networks from different classes with the identical set of parameters, namely with the number of nodes $N=750$ and the link probability $p=0.08$.  Solid curves in \fig{fig:04} (black for random Erd\H{o}s-Renyi (ER) network, red for e-network and blue for i-network) represent mean distributions, averaged over $n = 1000$ simulations on each networks, while shadowed regions designate standard deviations. The parameters $\beta$ and $\gamma$ are set to $\beta =0.05$, $\gamma=0.03$. The numerical results are reproduced for different network realizations.

Analyzing distributions in \fig{fig:04}, let us point out two important features of the epidemic spreading described by SIR model on different network archetypes. Apart from the maximal distribution on non-clustered ER network (black curve) which is our reference state, the interesting features demonstrate e-networks and i-networks. It turns out that clustering actually weakens the epidemic spread, but details are very sensitive to the way how the clustered network is constructed. The evolutionary grown e-networks demonstrate better suppression of epidemic spreading than instantly created i-networks for the same set of parameters. Meanwhile, the peak of the distribution of infected agents on e-networks is shifted to later times compared to both random and i-networks.

\section{Results and discussion}

Here we briefly speculate about a rather provocative title of our work in the context of selection between two different protocols of suppressing virus distribution. Currently two main mechanisms of a human population clustering are exploited by different countries in order to prevent the uncontrolled spreading of COVID-19. Tentatively these mechanisms could be named a "self-isolation" (SI) and a "frontiers closing" (FC). In both cases the aim of clustering is to localize the illness in closed community and prevent it from the propagation through the entire human network. Specifically, we are interested in the question which mechanism blocks better an epidemic spreading: self-quarantine in local communities induced by adaptive clustering, or sharp clustering via closing of borders between cities and countries? In an ideal situation, when all self-isolated communities are absolutely disconnected from each other, and when the border crossings between cities and countries is totally prohibited, both protocols are equally efficient and definitely inhibit disease expansion. However, in reality, it is impossible to isolate communities completely and some fraction of cross-community connections always is present. Readers are invited to make their own judgement whether such a speculation seems plausible and to which extent.

We have demonstrated that the network which is evolutionary grown from a randomly generated Erd\H{o}s-R\'{e}nyi graph with fixed vertex degree under condition of maximization of small cliques (triadic motifs) gets clustered into communities-clusters and the number of such communities depends on the linking probability $p$ in the initial graph (see \eq{eq:02}). We have also verified that similar adaptive clustering occurs when triadic motifs are replaced by complete 4-cliques. Running SIR model on e-networks, and in parallel, on i-networks (which mimic clustered structure of e-networks, however are memory-less), we see from \fig{fig:04} that e-networks prevent better epidemic spreading than i-networks (the maximum of infected agents is lower for e-networks), while the maximum of infected agents is shifted to later times compared to i-networks.

Importantly, we have found that the clustered e-network are scale-free. That explains some previous numerical observation concerning spectral density of such adaptively grown networks. We have also proposed new mechanism of derivation of scale-free behavior via peculiar rewiring process. Epidemic spreading on scale-free network has some specific peculiarities \cite{scale-epidemy}. In particular, the epidemic threshold almost vanishes which means that scale-free network is bad for epidemic suppression at the beginning of its distribution. However once started, it can be operated on scale-free network more effectively than on other types of networks. The new rewiring mechanism for getting scale-free behavior can be useful for these purposes.

We are grateful to O. Yartseva for pushing us to think about the impact of the underlying network structure on epidemic spread and to M. Tamm for valuable discussions. The work of OV is supported within the framework of the Basic Research Program at the NRU Higher School of Economics in 2020. SN and AG acknowledge the supports of foundations BASIS (17-11-122-1 for AG and 19-1-1-48-1 for SN), and RFBR (18-29-13013). The work of VA is supported within frameworks of the state task for the FRC CP RAS \# FFZE-2019-0016.


\begin{thebibliography}{99}

\bibitem{review} Pastor-Satorras, R., Castellano, C., Van Mieghem, P. and Vespignani, A. Epidemic processes in complex networks. Rev. Mod. Phys. 87, 925–979 (2015)

\bibitem{clust1} Serrano, M. Á. and Bogun\'{a}, M. Percolation and epidemic thresholds in clustered networks. Physical Review Letters 97, 088701 (2006)

\bibitem{clust2} Keeling, M. J. (1999), Proc. R. Soc. Lond. B 266, 859.

\bibitem{clust3} Moore, C., and M. E. J. Newman (2000), Phys. Rev. E 61, 5678.

\bibitem{adapt1} Gross, T., C. D’Lima, and B. Blasius (2006), Physical Review Letters 96 (20), 208701

\bibitem{anderson} Avetisov, V., Gorsky, A., Nechaev, S., and Valba, O. Localization and non-ergodicity in clustered random networks, Journal of Complex Networks, (2019) CNZ026, {\tt https://doi.org/10.1093/comnet/cnz026}, e-Print: arXiv:1611.08531

\bibitem{loc1} P. Sala, T. Rakovszky, R. Verresen, M. Knap, and F. Pollmann, arXiv:1904.04266 (2019)

\bibitem{loc2} V. Khemani and R. Nandkishore, arXiv:1904.04815 (2019)

\bibitem{loc3} F. Pietracaprina and N. Laflorencie, arXiv:1906.05709 (2019)

\bibitem{fortunato10} Fortunato, S. Community detection in graphs. Physics Reports 486.3-5, 75-174 (2010)

\bibitem{decelle13} Decelle, A., et al. Asymptotic analysis of the stochastic block model for modular networks and its algorithmic applications. Physical Review E 84.6, 066106 (2011)

\bibitem{vir} G\'{a}bor Vattay, Predicting the ultimate outcome of the COVID-19 outbreak in Italy, arXiv:2003.07912

\bibitem{maslov} S. Maslov and K. Sneppen, Specificity and Stability in Topology of Protein Networks, Science, 296, 910 (2002)

\bibitem{schelling} V. Avetisov, A. Gorsky, S. Maslov, S. Nechaev, and O. Valba, Social behavior beyond the Schelling model, Phys. Rev. E 002300 (2018)

\bibitem{hovan} Avetisov, V., Hovhannisyan, M., Gorsky, A., Nechaev, S., Tamm, M., and Valba, O. Eigenvalue tunneling and decay of quenched random network. Phys. Rev. E 94, 062313 (2016)

\bibitem{reconnection} F. Viger and M. Latapy, Efficient and simple generation of random simple connected graphs with prescribed degree sequence,  Computing and Combinatorics, Lecture Notes in Computer Science 3595, 440 (2005)

\bibitem{bianconi} Hamed Seyed-allaei, Ginestra Bianconi, Matteo Marsili, Scale-free networks with an exponent less than two, Phys. Rev. E 73, 046113 (2006)

\bibitem{dorog} Gábor Timár, Sergey N. Dorogovtsev, José Fernando F. Mendes, Scale-free networks with exponent one, Phys. Rev. E 94, 022302 (2016)

\bibitem{sir} Hethcote H. The Mathematics of Infectious Diseases. SIAM Review. 42 (4): 599–653 (2000)

\bibitem{scale1} Illes J. Farkas, Imre Derenyi, Albert-Laszlo Barabasi, Tamas Vicsek, Spectra of "Real-World" Graphs: Beyond the Semi-Circle Law, Phys. Rev. E 64, 026704:1-12 (2001)

\bibitem{scale2} K.-I. Goh, B. Kahng, D. Kim, Spectra and eigenvectors of scale-free networks, Phys. Rev. E 64, 051903 (2001)

\bibitem{scale-epidemy} Pastor-Satorras, R. and Vespignani, A. Epidemic spreading in scale-free networks. Phys. Rev. Lett. 86, 3200 (2001).

\end{thebibliography}
\end{document}